\documentclass[10pt,journal,compsoc]{IEEEtran}

\ifCLASSOPTIONcompsoc
  \usepackage[nocompress]{cite}
\else
  \usepackage{cite}
\fi
\usepackage{tabularx}
\usepackage{multirow}
\usepackage{balance}
\usepackage{graphicx}
\usepackage{textcomp}
\usepackage{xcolor}
\def\BibTeX{{\rm B\kern-.05em{\sc i\kern-.025em b}\kern-.08em
    T\kern-.1667em\lower.7ex\hbox{E}\kern-.125emX}}
    
\ifCLASSINFOpdf
\else
\fi
\usepackage{amsmath,amssymb,amsfonts}
\usepackage{algpseudocode}
\usepackage{algorithm}

\usepackage{pgfplots}

\begin{document}
%
\title{\MechanismNoSpace: Decentralized, Scalable, and Consistent Storage for the Edge}
%
%
%
%

\author{\textbf{Karim Sonbol},
        \textbf{\"Oznur \"Ozkasap},
        ~\textbf{Ibrahim Al-Oqily},~\textbf{Moayad Aloqaily}
        \IEEEcompsocitemizethanks{
\IEEEcompsocthanksitem Karim Sonbol and  \"Oznur \"Ozkasap are with Department of Computer Engineering, Ko\c{c} University, Istanbul, Turkey. \protect E-mails: \{ksonbol16; oozkasap\}@ku.edu.tr

\IEEEcompsocthanksitem Ibrahim Al-Oqily is with The Hashemite University, Zarqa, Jordan and Al-Yamamah University, Riyadh, Saudi Arabia. \protect E-mail: ialoqily@ieee.org

\IEEEcompsocthanksitem Moayad Aloqaily is with Al Ain University, UAE. \protect E-mail: moayad@gnowit.com
}}

\newcommand{\Mechanism}{EdgeKV\ }
\newcommand{\MechanismNoSpace}{EdgeKV}

\IEEEtitleabstractindextext{%
\begin{abstract}
Edge computing moves the computation closer to the data and the data closer to the user to overcome the high latency communication of cloud computing. Storage at the edge allows data access with high speeds that enable latency-sensitive applications in areas such as autonomous driving and smart grid. However, several distributed services are typically designed for the cloud and building an efficient edge-enabled storage system is challenging because of the distributed and heterogeneous nature of the edge and its limited resources.
In this paper, we propose \MechanismNoSpace, a decentralized storage system designed for the network edge. \Mechanism offers fast and reliable storage, utilizing data replication with strong consistency guarantees. With a location-transparent and interface-based design,
\Mechanism can scale with a heterogeneous system of edge nodes. 
We implement a prototype of the \Mechanism modules in Golang and evaluate it in both the edge and cloud settings on the Grid'5000 testbed. We utilize the Yahoo! Cloud Serving Benchmark (YCSB) to analyze the system's performance under realistic workloads. Our evaluation results show that \Mechanism outperforms the cloud storage setting with both local and global data access with an average write response time and throughput improvements of 26\% and 19\% respectively under the same settings. Our evaluations also show that \Mechanism can scale with the number of clients, without sacrificing performance. Finally, we discuss the energy efficiency improvement when utilizing edge resources with \Mechanism instead of a centralized cloud.
\end{abstract}

}

\maketitle

\IEEEdisplaynontitleabstractindextext

%
\IEEEpeerreviewmaketitle

\ifCLASSOPTIONcompsoc
\IEEEraisesectionheading{\section{Introduction}\label{sec:introduction}}
\else
\section{Introduction}
\label{sec:introduction}
\fi

Several distributed services need to store their state and change it or retrieve it at a later time. This state is generally required to be stored in a reliable, secure, private, efficient, and cost-effective way. The cloud has been traditionally used for storing the state of many programs. The cloud provides flexible pay-what-you-use cost policy, scalability, high security standards, and high reliability. The cloud is able to provide such features by hosting clusters of commodity servers in one physically secure location; namely a data center \cite{datacenter-virtual}.

Data centers are sparse because of their special requirements and high costs for building and maintenance. This implies that the majority of cloud users would be distant from data centers. Such distances are important because they directly affect the latency of data transfer between clients and servers, which becomes more critical in latency-sensitive systems such as autonomous vehicles, smart grids, and online multiplayer games. Another problem with cloud storage is data privacy since sending data to the cloud typically means sharing it with a third-party. Hence, applications storing sensitive information may need to find an alternative to the cloud.

To overcome the high latency of the cloud, fields such as edge, fog, and mist computing came into existence \cite{balasubramanian2019reinforcing}\cite{edge-of-things}. In this paper, we refer to these entities collectively as edge nodes. These fields aim to move the data and computation closer to the consumer. Therefore, instead of doing most of the computation and storage at the cloud, new computation and storage entities can be introduced between the client and data center in the client-to-cloud continuum to partially handle these tasks. This not only reduces the response latency and saves network bandwidth but also alleviates some of the workload from the cloud. In addition, storing and processing data at the edge allow utilizing contextual information to improve data locality and decision making \cite{edge-iot}. Moreover, privacy-oriented applications can store their data on the local edge, or use the cloud for storing only less-sensitive or aggregate information. 

Edge nodes generally have limited computation, storage, network, or power resources. They may have heterogeneous hardware and software architectures, and are not typically located in one location. Therefore, they may communicate over a Wide-Area Network (WAN) and some even use a wireless protocol, e.g., WiFi or 4G/5G, to communicate together or with clients. Consequently, designing an edge-enabled storage system is not trivial
and directly using existing cloud-optimized systems in the edge may not be possible. Instead, either a middleware layer can be introduced to make the existing software edge-aware or novel designs can be developed that are specific to the edge or that are generic enough to work with both cloud and edge.

A few existing works try to build an edge-capable storage system but they are either domain-specific \cite{vision,openstack-edge}, provide only weak forms of consistency \cite{eclipse-fog}, or require a high level of domain knowledge and customization to work efficiently \cite{fogstore}. 
In this paper, we propose \MechanismNoSpace, a decentralized, general-purpose, scalable, and reliable storage system for the edge. \Mechanism offers low-latency access, strong consistency notions, high availability, and minimal customization requirement.
Specifically, we make the following contributions:
\begin{itemize}

\item Propose \MechanismNoSpace, a novel storage system architecture for the edge, explain its modules and their interactions, and the algorithms used. 

\item Provide fault-tolerance and reliability through replication in the edge with strong consistency guarantees. Support load-balancing with a highly-scalable overlay that has minimal overhead. 

\item Provide of two levels of data locality and privacy to provide differential latency guarantees based on application requirements. The separation of local and global data allows deploying \Mechanism in different use cases.

\item Develop a prototype \cite{edgekv-github} allowing for heterogeneity in the system with a hierarchical design and interface abstractions. Perform comprehensive experimental analysis on distributed testbed and comparisons with the centralized cloud storage.

\item Present the performance analysis results of \MechanismNoSpace, showing its superior performance to the centralized cloud solution, especially with local data, and evaluate the scalability of the system with the number of clients and requests. In addition, we provide a discussion about the energy efficiency aspect of the system. Moreover, we discuss possible optimizations for scalability, future research directions for \MechanismNoSpace, and useful insights for edge-enabled application designers.

\end{itemize}

The rest of the paper is organized as follows. We discuss the related work in section \ref{sec-rel-work}. Section \ref{sec-sys-arch} introduces the design and system architecture. Section \ref{sec-use-case} discusses motivating use cases for \MechanismNoSpace. In section \ref{section:emulation-setup}, we discuss implementation details and in section \ref{sec-perf-resu} we present evaluation results. Finally, we conclude in section \ref{sec-concl} with a summary of the contributions and future directions.

\section{Related Work}\label{sec-rel-work}
There exists a variety of work on distributed storage for the cloud. This ranges from relational databases such as MySQL and PostgreSQL to NoSQL databases such as Cassandra and MongoDB and includes key-value stores such as TiKV. However, less work exists for utilizing the edge/fog resources. Nonetheless, we discuss the relevant existing works next, also summarized in table \ref{table:related-work}. 

FogStore \cite{fogstore} is designed for situation-awareness applications that use data annotated with context information such as location or timestamps. FogStore presents a geo-replicated key-value storage providing differential consistency guarantees based on the context. While this allows for usage in different scenarios, it requires the system user to have expert knowledge of the domain to define a mapping between data and client contexts and the required consistency level for different queries. Similarly, Vision-Edge \cite{vision} is an application-specific key-value storage solution for machine vision applications such as smart surveillance cameras. There are two types of data in such applications: latency-critical feature vectors and key-frames stored for bookkeeping purposes. FBase \cite{fbase} is a replication service for data-intensive fog applications. It provides programmers with a declarative way to choose replication paths and data flows across geo-distributed sites, based on user-provided configuration data. FBase guarantees only eventual consistency for the application data and strong consistency for configuration data.  
Some works do not handle fault-tolerance such as Vision-Edge and Edge-Cloud+ (EC+) \cite{mmog-edge}, while others provide fault-tolerance but with weak forms of consistency such as Fog05 and Workers-KV \cite{workers-kv}. EC+ is an architecture augmented by edge computing for Massively Multiplayer Online Games with Virtual Reality (VR-MMOG). EC+ utilizes the edge for latency-sensitive local view change updates and leaves global game state updates, with less strict  latency requirements, to the central cloud. CloudFlare's Workers-KV \cite{workers-kv} utilizes CloudFlare's global edge network to build low-latency globally-available key-value storage. Workers-KV only provides eventual consistency and is mainly useful for building faster and customized web applications

We note that each work has a specific use case or class of use cases for which they are designed. For example, Fog05 \cite{eclipse-fog} and \cite{revising-openstack} are Infrastructure-as-a-Service (IaaS) frameworks that are used for managing both cloud and edge resources. Eclipse fog05 \cite{eclipse-fog} is a virtualization solution for cloud, edge, and fog resources suitable for heterogeneous systems. It can integrate any key-value store and provide a location-transparent and unified view to it from anywhere in the network through a unified interface, but provides only eventual consistency. OpenStack-Edge modifies the OpenStack IaaS framework to enable managing edge resources. It achieves this by replacing its centralized SQL database with a distributed Redis cluster. Dqlite \cite{dqlite} provides a distributed, highly available, and lightweight SQLite implementation suitable for embedded devices.
While Dqlite provides strong consistency and fault-tolerance, it designed to be used only with a small number of servers. Finally, the proposed \Mechanism system is a general-purpose system that provides fault-tolerance with strong consistency guarantees for storage of key-value pairs.

\begin{table*}
\begin{center}
\caption{\label{table:related-work} Summary of existing edge-enabled storage solutions.}
\bgroup
\def\arraystretch{1.4}
\begin{tabularx}{0.9\textwidth}
{
| >{\centering\arraybackslash}X
| >{\centering\arraybackslash}X
| >{\centering\arraybackslash}X
| >{\centering\arraybackslash}X
| >{\centering\arraybackslash}X
|
}
\hline
Project & Use Case & Data Content & Consistency & Fault-tolerance\\
\hline
FogStore \cite{fogstore}& situation-awareness, & \multirow{2}{*}{contextual data} & \multirow{2}{*}{context-based} &
\multirow{2}{*}{Yes}\\
& applications &&&\\
\hline
Vision-Edge \cite{vision}& \multirow{2}{*}{Computer vision} & feature vectors, & \multirow{2}{*}{Data-type-based} & \multirow{2}{*}{No}  \\ 
 &&key-frames&&\\
\hline
FBase \cite{fbase}& Data-intensive fog & Application data & Eventual & \multirow{2}{*}{Yes}\\
\cline{3-4}
 &applications& Configuration data & Strong &\\
\hline
EC+ \cite{mmog-edge}&\multirow{2}{*}{MMOG}&\multirow{2}{*}{Game events}&\multirow{2}{*}{Event-type-based} & \multirow{2}{*}{No}\\
&&&&\\
\hline
Workers-KV \cite{workers-kv}& \multirow{2}{*}{Web services} & \multirow{2}{*}{Web pages} & \multirow{2}{*}{Eventual} & \multirow{2}{*}{Yes} \\ 
&&&&\\
\hline
Fog05 \cite{eclipse-fog}& \multirow{2}{*}{IaaS} & \multirow{2}{*}{Server states} & \multirow{2}{*}{Eventual} & \multirow{2}{*}{No}\\ 
&&&&\\
\hline
OpenStack-Edge \cite{revising-openstack}& \multirow{2}{*}{IaaS} & \multirow{2}{*}{Server states} & \multirow{2}{*}{Eventual} & \multirow{2}{*}{Yes}\\ 
&&&&\\
\hline
Dqlite \cite{dqlite}& \multirow{2}{*}{Embedded devices} & \multirow{2}{*}{Sensor data} & \multirow{2}{*}{Strong} & \multirow{2}{*}{Yes} \\ 
&&&&\\
\hline
EdgeKV & General-purpose & Key-Value pairs & Strong & Yes\\
\hline
\end{tabularx}
\egroup
\end{center}
\end{table*}

\section{System Architecture}\label{sec-sys-arch}
\Mechanism offers a decentralized storage architecture for the edge with strong consistency guarantees through state machine replication. \Mechanism connects independent edge groups with a ring overlay for high scalability. We explain the system architecture of \Mechanism in this section and  present both a layered and modular view of the system. Table  \ref{table:notations} summarizes the notation used throughout the paper.
\begin{table}[t]
\begin{center}
 \caption{\label{table:notations} Summary of the notations used and their definitions.} \bgroup
\def\arraystretch{1.3}
\begin{tabularx}{0.4\textwidth}
{
| >{\centering\arraybackslash\hsize=0.2\hsize}X
| >{\arraybackslash\hsize=0.8\hsize}X
|
}
\hline
Notation & Definition\\
\hline
n & number of nodes in an edge group\\
\hline
m & number of gateway nodes in the system\\
\hline
G & total number of global keys in the system\\
\hline
L & number of keys in a local edge group\\
\hline
S & average size of a local key-value pair\\
\hline
T & average size of a global key-value pair\\
\hline
Cli & client \\
\hline
Gw & gateway node\\
\hline
St & storage node\\
\hline
\end{tabularx}
\egroup
\end{center}
\end{table}
\subsection{Layered view}
\begin{figure}[h]
\includegraphics[scale=0.7,width=7cm]{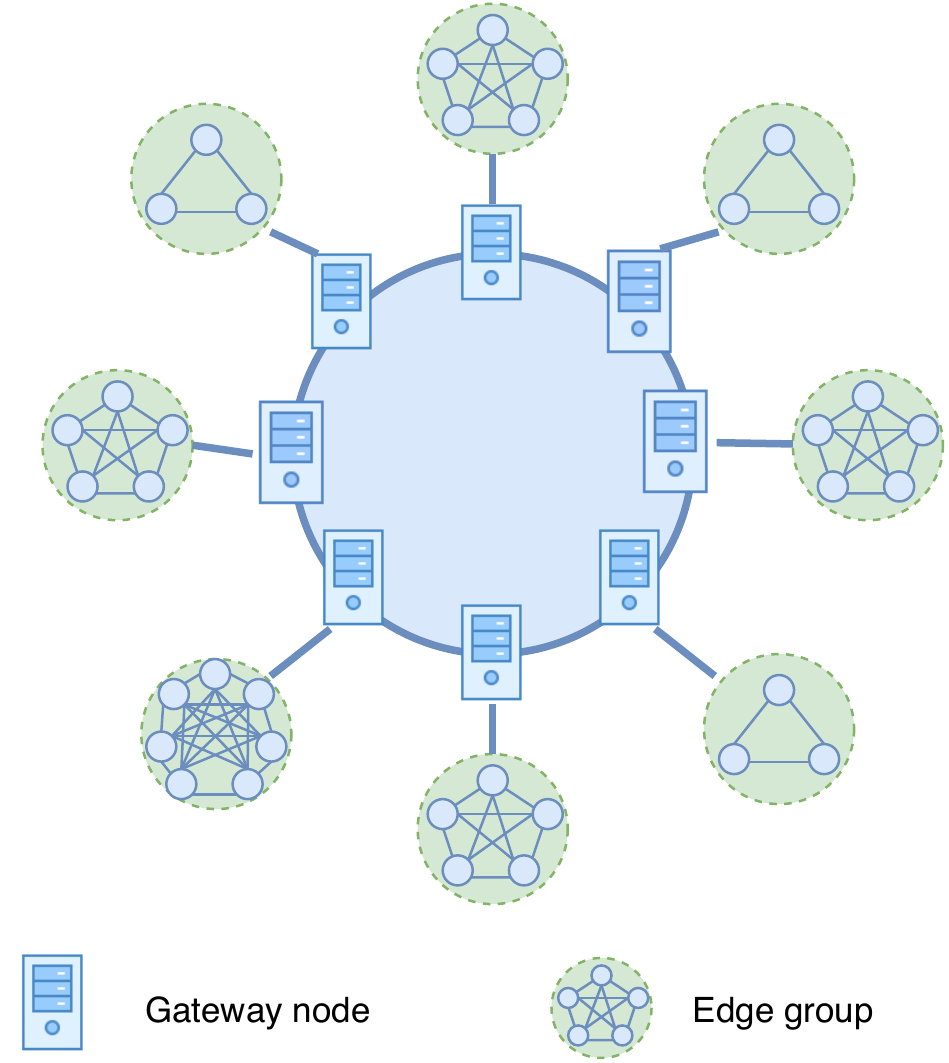}
\centering
\caption{System architecture including both the local groups formed of nearby edge nodes and the overlay of gateway nodes.}
\label{fig:dht_groups}
\end{figure}

\Mechanism design adapts a hierarchical approach, building from a small number of local nodes to a large-scale system. \Mechanism is composed of two layers: a local layer of independent groups of edge servers and a global layer connecting these groups through a ring overlay, as shown in Fig.\ref{fig:dht_groups}.

\textbf{In the local layer}, a small number of edge nodes which are located in a close geographical proximity form a group. This group represents a replicated state machine (RSM). In other words, each node in this group has a copy of the same state (i.e., key-value pairs) for fault-tolerance.
However, a write operation to the edge group is considered complete once a majority of its members have replicated the data item. This allows the edge group to function correctly even with the failure of a minority of its members.
A consensus protocol maintains strong consistency between the group members. This means that (concurrent) read and write requests will be applied to all members of the state machine in the same order and no stale values would be returned to a user at any time. These local groups are independent, so they can employ different characteristics such as different group sizes (as shown in Fig. \ref{fig:dht_groups}) or different replication strategies.

\textbf{The global layer} is the upper layer in the hierarchy. In a typical deployment of \MechanismNoSpace, there would be many groups spread over a large geographical area (e.g., a city or a country). These groups together form an overlay or a global layer that allows for scalability and the two layers together constitutes the system. In the global layer, different groups are connected through \textit{gateway nodes}. A single gateway node is located close to at least one of the system groups and is responsible for forwarding data from that group to any remote group, and vice versa. It achieves this by first locating the remote group’s gateway node in the overlay and then sending data to or asking for data from that group. 
Since different edge groups communicate only through the gateway nodes, this provides great flexibility and room for heterogeneity. Different edge groups can have different sizes, can use different internal replication mechanisms, and possibly build on different hardware and software architectures.

\textbf{A distributed hash table (DHT)} forms the global overlay of gateway nodes. A DHT ensures a fair distribution of keys from the key-space among the gateway nodes, hence their corresponding groups. In a DHT, each node is given a unique hash based on its identifier (e.g., IP address) which specifies its location in the overlay. Utilizing a consistent hashing scheme ensures that such hashes are fairly distributed along the overlay and are collision-resistant. Similarly for key-value pairs, the hash of the key decides its location, hence its responsible node in the overlay. In \Mechanism design, gateway nodes are used only for routing a key-value pair to its corresponding group. The key-value pairs are stored and replicated in the local edge groups. The gateway node itself stores only routing information needed for the DHT. For n gateway nodes, DHT requires O(log(n)) storage complexity on each node to achieve O(log(n)) message complexity for locating any node in the overlay from any other node. Efficient routing and fair load distribution in a DHT allows for scalability.

\subsection{Modular view}
\begin{figure}
\includegraphics[scale=0.25,width=7cm]{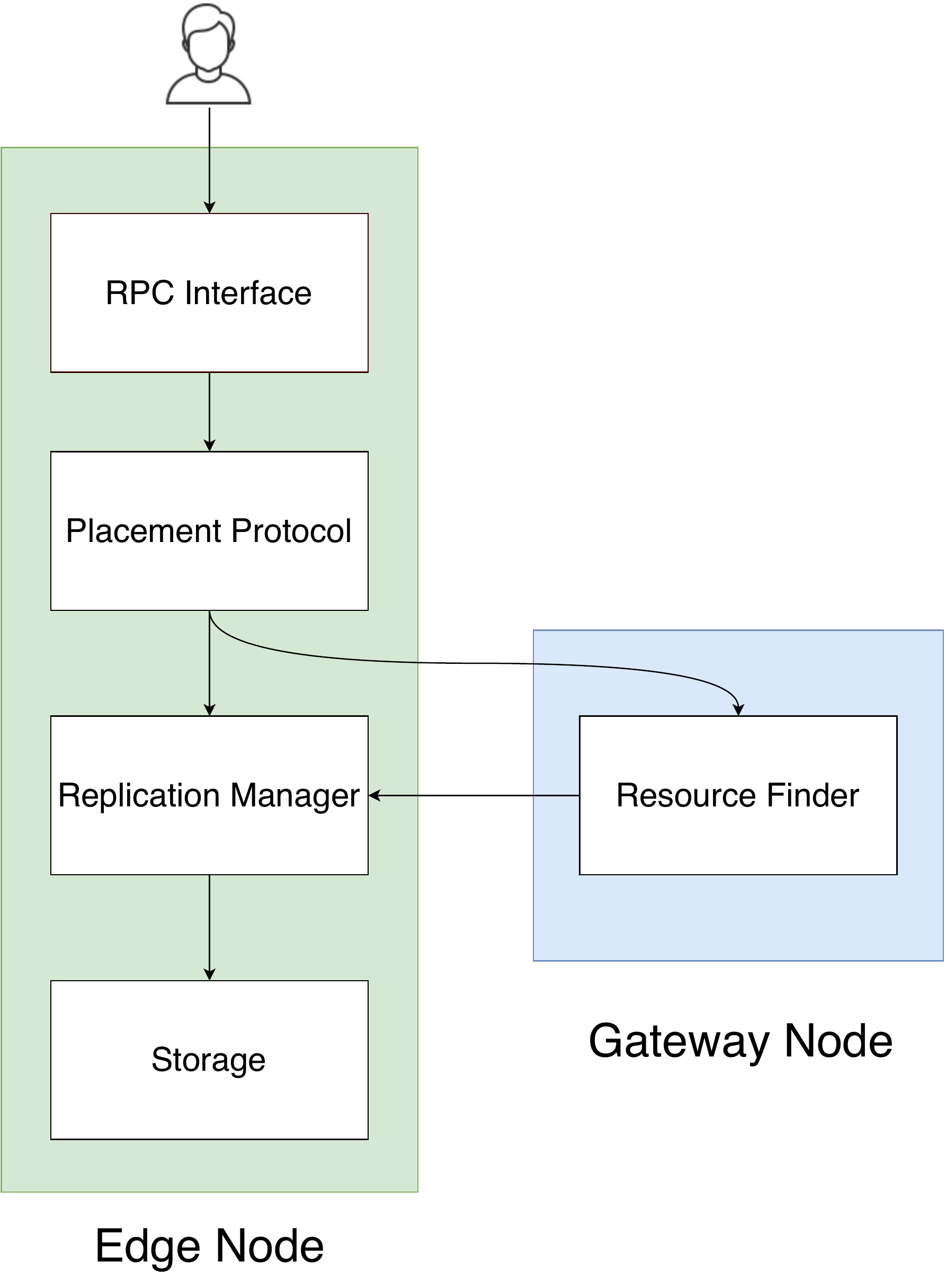}
\centering
\caption{EdgeKV modules and their interactions, separated into edge node modules and gateway node module}
\label{fig:modules}
\end{figure}

The discussed layers are implemented with modules running on the edge and gateway nodes. \Mechanism comprises the modules of Remote Procedure Call (RPC) interface, placement protocol, resource finder, replication manager, and storage. These modules and their interactions are shown in Fig. \ref{fig:modules}. The proposed application flow is as follows: a client communicates with its closest edge node through the RPC interface, specifying the type of operation (e.g., get, put, and delete), the key (and possibly the value) to perform the operation on, and the data type (i.e., local vs global). Then, the placement protocol (on the edge node) decides, based on the data type, whether to perform the operation in the current edge group or to forward it to a remote group through the resource finder. The resource finder (on the gateway node) utilizes the DHT overlay to decide which edge group is responsible for that key. Afterwards, it forwards the request to that group through its assigned gateway node. Then, the assigned edge group performs the operation on all group members through the replication manager through a consensus quorum. Finally, the storage module handles the actual storage, retrieval, or deletion of data on the physical storage media. In the following sections, we describe each module in detail with its algorithm.

\subsubsection{RPC interface} is provided to the end nodes to allow them to store or access the data. The interface supports GET, PUT, and DELETE operations for key-value pairs, where PUT is used for creating a new key-value pair or updating the value of an existing key. Utilizing an RPC interface allows the system to change its internal implementation without changing the users' code. Moreover, the RPC module provides a structured and efficient method of communication.

\subsubsection{Placement Protocol} on the edge nodes decides where data should be placed based on its type: local or global. As depicted in Algorithm \ref{alg:placement}, local data are stored in the local storage group whereas global data are distributed over the system groups. The placement protocol forwards local data to the replication manager on the same edge node and global data to the resource finder on the local gateway node. 
\begin{algorithm}
    \caption{Placement Protocol}\label{alg:placement}
    \begin{algorithmic}[1]
        \Function{$placement$}{$key, value, type$}
            \If{$type == local$}
                \If{$nodeType == Leader$}
                    \State \Call{$replicate$}{key, value}
                \Else
                    \State \Call{$send$}{Leader, key, value, type}
                \EndIf
            \Else
                \State \Call{$send$}{gateway, key, value}
            \EndIf
        \EndFunction
    \end{algorithmic}
\end{algorithm}
\subsubsection{Resource Finder} runs on the gateway nodes and utilizes the DHT overlay to decide in which edge group a key-value pair should be stored and the location of such group in the overlay. When the resource finder receives a request from its local edge group for global data access or storage, it hashes the key and decides which edge group is responsible for that key. Then, it routes the request to the gateway node associated with that group using standard DHT routing.
Once the remote gateway node receives the request, it forwards it to the replication manager in its assigned edge group. Pseudo-codes for the get and put functions of the resource finder are shown in Algorithm \ref{alg:finder}.

\begin{algorithm}
    \caption{Resource Finder}\label{alg:finder}
    \begin{algorithmic}[1]
        \Function{$put$}{$key, value$}
            \State  $keyHash \gets \Call{$hash$}{key}$
            \State $targetNode \gets overlay. \Call{$locate$}{keyHash}$
            \State $response \gets targetNode. \Call{$replicate$}{key, value}$
            \State \Return{$response$}
        \EndFunction
        \Statex
        \Function{$get$}{$key$}
            \State  $keyHash \gets \Call{$hash$}{key}$
            \State $targetNode \gets overlay. \Call{$locate$}{keyHash}$
            \State $value \gets targetNode. \Call{$get$}{key}$
            \State \Return{$value$}
        \EndFunction
    \end{algorithmic}
\end{algorithm}

\subsubsection{Replication Manager} 
The replication manager runs on each edge node in an edge group and is responsible for replicating data to all the of group nodes. It uses a consensus protocol to ensure strong consistency among the edge nodes. The consensus protocol elects a leader from the group that handles the replication of all group data during its term. The leader may change dynamically if the current leader fails or becomes overloaded. The replication manager is responsible for the consistency of read and write operations to the storage module.

We select the Raft protocol \cite{raft-extended} as the consensus protocol used by the replication manager module. Raft is a leader-based consensus protocol providing strong consistency with a simple and understandable design. We explain below the two main processes of Raft, namely the leader election and append entries processes.

\subsubsection{Storage} module handles the actual storage of the key-value pairs on each edge node. Two separate key-value stores are available on each node, a local one for group-level data, and a global one for system-level data. An end node has access to the global storage and the local storage of its connected group only.

\section{Use Cases}\label{sec-use-case}

In this section, we discuss potential use cases for our edge-optimized storage system. However, \Mechanism is a general-purpose system suitable for many other uses, especially where data consistency and low-latency data access is significant. We discuss two such use cases, namely smart grid and autonomous driving which can benefit from the low latency of the edge to achieve their requirements. The edge can be utilized for low-latency data access while the cloud can still be used for data of larger sizes and lower latency requirements.

\subsection{Smart Grid}
The smart grid (SG) allows real-time monitoring and prediction of consumer loads and the integration of renewable energy sources. By utilizing many sensors and learning models, SG allows energy suppliers to save energy by correctly predicting demand at peak hours without overestimating. SG also ensures a fair system by allowing suppliers to dynamically update the energy price in real-time to which consumer systems can adapt their loads \cite{edge-sc-chapter}.

Current SGs send all data to the cloud for storage and computation. However, the high-latency of the cloud may not be tolerated in some cases, and the large amount of data sent from an SG in real-time can overload the network, causing even more latency and performance degradation. 
Additionally, since an SG system often involves sharing private information (e.g., readings from a smart meter in a smart home), sending such data to the cloud makes it more possible to be exposed to unwanted third parties.

Edge/fog computing solves these issues by storing data closer to its sources and doing some accumulation, pre-processing, or filtering on data before sending it to the cloud. This way, less data is sent through the network, better response times are achieved, and consumer privacy is protected by sending only accumulated or non-sensitive data to the cloud. 
An SG can specifically utilize our system for storing energy usage information (e.g., readings from a smart meter) on the edge. Energy suppliers can also communicate with each other and with the consumers through our decentralized overlay.

\subsection{Autonomous Driving}
Autonomous driving aims to make transportation safer, more organized, and energy-efficient. An autonomous vehicle uses its on-board sensing devices (e.g., cameras, radars, proximity sensors) to understand its close surroundings. Additionally, it uses information from other vehicles, edge, and cloud infrastructures to understand its out-of-sight environment and learn useful information such as high definition maps, traffic lights status, and traffic conditions. This allows for dynamic path planning and having safer and more efficient trips. The decisions made by an autonomous vehicle are often complex and latency-critical and the cost of a mistake can be very high. Thus, the low latency data access provided by the edge is in many cases preferred to the high latency of long-haul communication with the cloud. The edge can be used for storage and doing critical computations.

Specifically, autonomous vehicles may utilize our system for crowdsourcing local traffic information and gaining low-latency access to information needed for making critical navigation decisions. The cloud can still be used for gaining access to less critical information such as the closest attractions or expected weather forecast.


\section{Platform Setup} \label{section:emulation-setup}
In this section, we discuss the tools used for the prototype implementation and explain our emulation setup and the evaluation framework we used for performing different performance benchmarks of \MechanismNoSpace. We also discuss the performance metrics used and their significance in the evaluation.

\subsection{Prototype Implementation} \label{sec-impl}
\begin{figure}[h]
\includegraphics[scale=0.3]{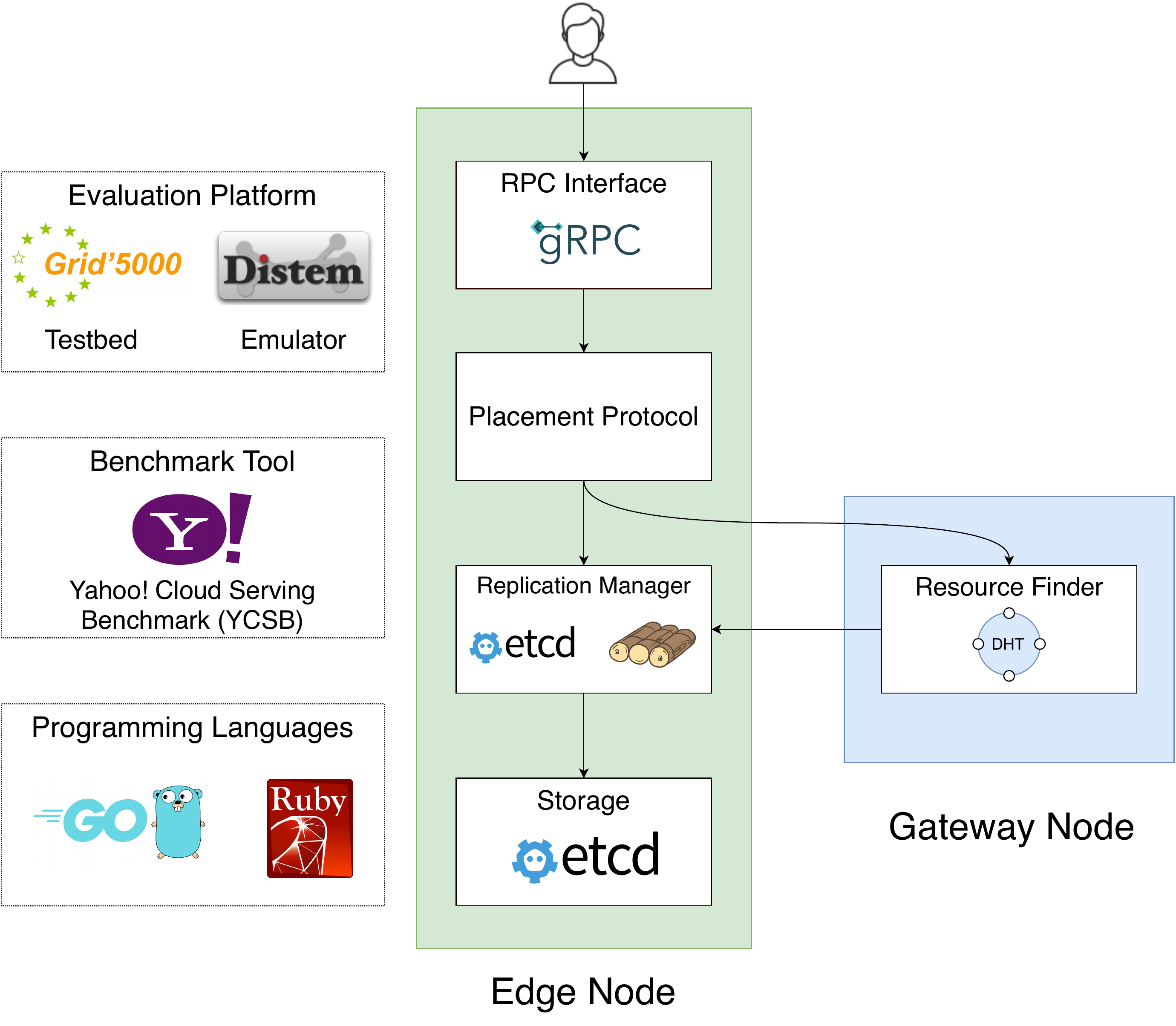}
\centering
\caption{The tools used for the implementation and evaluation of \MechanismNoSpace.} \label{fig:tools}
\end{figure}

We developed a prototype of the proposed \Mechanism edge storage system in the Go programming language. The tools used for implementation and evaluation of \Mechanism are summarized in Fig. \ref{fig:tools}. 

Each entity in our system, namely clients, edge nodes, and gateway nodes, runs a separate application. A client sends requests to an edge node which communicates with other edge nodes in the same group and, if needed, with the gateway node assigned to its group to fulfil the client request. Then, the edge node sends a response to the client after achieving the read or write consensus. Gateway nodes communicate with each other according to the DHT protocol to fulfil requests from edge nodes. Communication between the different system modules is achieved through gRPC interfaces which allow a simple, structured and efficient means of communication. gRPC is an open-source protocol designed by Google to be scalable, inter-operable, and general purpose.

The replication manager and storage modules are implemented using \textit{etcd}, a popular open-source key-value store used in many cloud deployments. etcd uses the Raft consensus protocol \cite{raft-extended} to ensure a strongly-consistent state between all replicas in an edge group. Raft is an efficient yet understandable leader-based consensus protocol widely used in RSMs. It starts by electing a leader from the group, then continues with data replication coordinated by the leader.

Because edge groups do not directly communicate, different consensus protocols and different storage backends may be used in different groups, as long as they implement the same gRPC interface and provide the same performance and consistency guarantees. 
For simplicity, we use the etcd key-value store for storage and replication management in all groups in our prototype.

The resource finder runs on the gateway nodes and utilizes a DHT overlay. We chose to implement the Chord DHT, one of the most widely used DHTs, following the optimized algorithms in \cite{chord}. Chord DHT has a simple design and maintains an O(log(n)) communication and storage complexity. DHT nodes also communicate through well-defined gRPC interfaces. Using the DHT allows us to create a highly-scalable key-value storage system using small independent etcd clusters.

\subsection{Evaluation Platform}
To evaluate the performance of our \Mechanism prototype, we use the Grid'5000 testbed with the Distem Network emulator to create the system setup we need. We also use the YCSB \cite{ycsb} tool for generating workloads and running a high number of operations against the system. The evaluation tools are summarized in Fig. \ref{fig:tools} and we explain each module in detail below.

\subsubsection{Grid'5000}
Grid'5000 is a popular large-scale testbed used by researchers for evaluating several kinds of systems with a focus on distributed and parallel systems, big data, Artificial Intelligence, and High-Performance Computing applications. It has 15,000 cores and 800 compute nodes organized in homogeneous clusters which are distributed over 8 sites in France.

We chose Grid'5000 as our testbed for a number of reasons. First, it provides bare-metal access to the servers which allows for great flexibility in setting up the needed software stack. Second, it provides a handful of useful tools for node reservation and deployment, and for experiment monitoring and result collection. Moreover, providing both a RESTful Application Programming Interface (API) and a Ruby client to the framework allowed automating most parts of the experiment process. Finally, an important advantage over other similar testbeds is the accuracy of node status data and the wide support available through the technical team and other users of the system.

\subsubsection{Distem}
While Grid'5000 provides flexible access to physical nodes with the desired compute, memory, network, and storage specifications, the Distem network emulator allows to build complex network topologies and setting up a large number of virtual nodes over a limited number of physical nodes in a short time. It also allows emulating different scenarios (e.g., cloud vs edge) by creating virtual networks with customizable link specifications, regardless of the underlying network architecture. Similar to Grid'5000, Distem also provides a command-line client, a REST API, and a Ruby client allowing to script the experiments for easy reproductability. In addition, it allows controlling the whole experiment from a single node, known as the coordinator node. 

\subsubsection{YCSB}
We use YCSB to generate realistic workloads and run them against our system. To integrate it with \MechanismNoSpace, we implemented a simple YCSB database interface layer that uses our \Mechanism client. YCSB generates realistic workloads, of which we choose the update-heavy workload "A" with 50\% read operations and 50\% write operations since our system performance gets affected by the percentage of write requests. YCSB also allows changing parameters such as the request distribution and data size. To simulate multiple concurrent connections, each client runs a 100 YCSB workers (i.e., threads) to send requests. In the next sections, we present different evaluations for different aspects of the system.

Experiments performed with YCSB have two phases. First, the \textit{load phase} when key-value pairs are inserted into the \Mechanism storage. Specifically, 10,000 key-value pairs are inserted into the edge nodes storage. Second, the \textit{run phase} when read and update operations are performed against the stored key-value pairs. To experiment with the concept of data types (i.e., local vs global), two changes have been made to the \Mechanism database interface layer: When creating the key-value pairs, we store two copies of each pair, one in the local storage, and the other in the global one. Next, requests are randomly chosen to be run on the local or global storage with a probability that is defined by the \textit{proportion of global data} parameter. This parameter is passed to YCSB at each benchmark run.

\begin{figure}[h]
\centering
\includegraphics[scale=0.7]{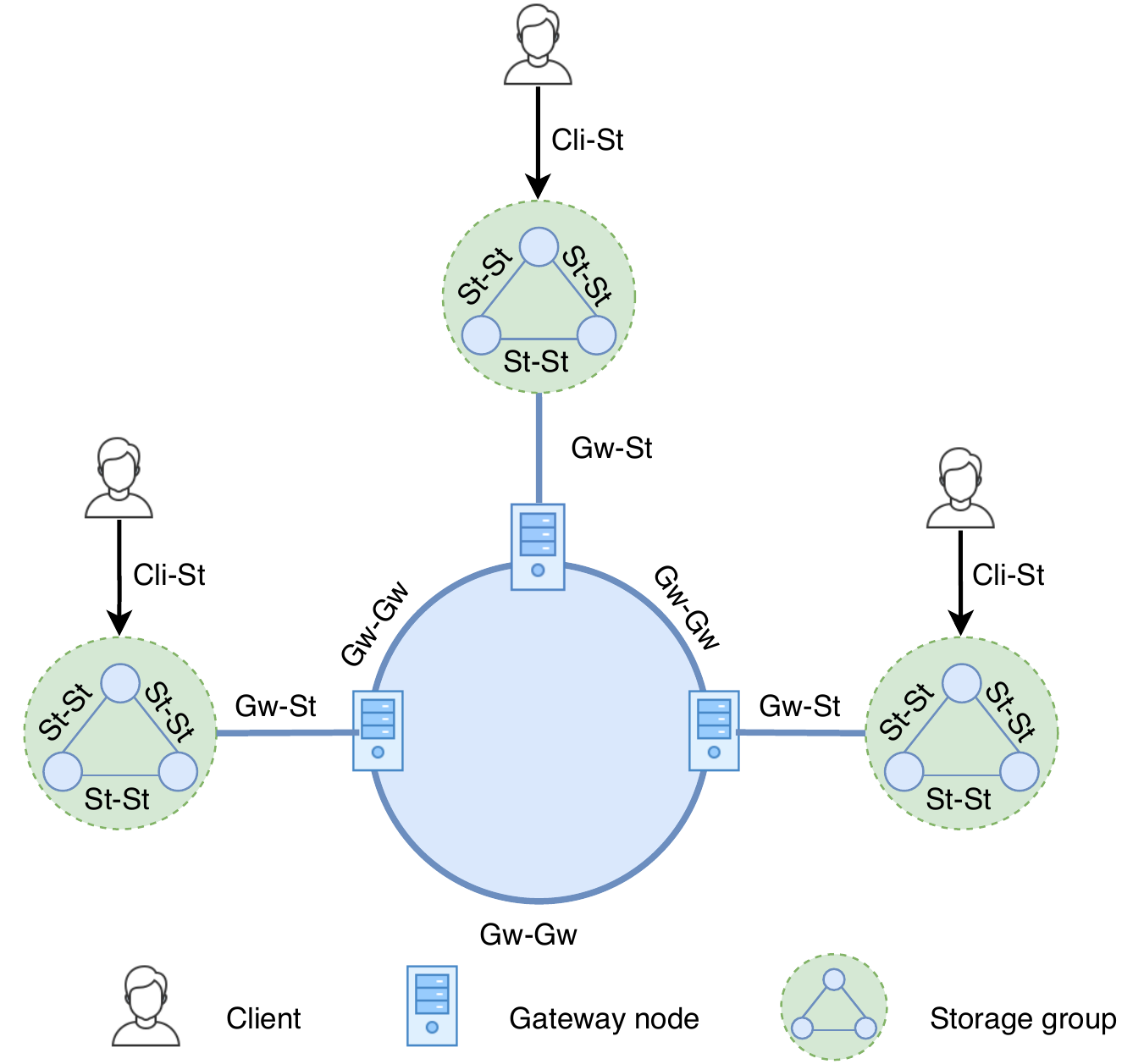}
\caption{The setup used for evaluating \MechanismNoSpace: three edge groups and their assigned gateway nodes and clients. Link labels are also shown with abbreviations: Cli: client, St: storage node, and Gw: gateway node.} \label{fig:eval-setup}
\end{figure}

\subsection{Experimental Setup}
Our evaluation setup is as follows: we set up three edge storage groups (the local layer), each consisting of three edge servers, and each is assigned one gateway node that is also a member in a DHT ring (the global layer). We also initiate three clients, one for each group, and each running 100 threads to simulate concurrent requests. The setup is shown in Fig. \ref{fig:eval-setup}.

Our setup has a total of 15 virtual nodes running on 15 physical machines in the 'grisou' cluster of the 'Nancy' site. Each node has 2 Intel Xeon E5-2630 v3 CPUs with 8 cores/CPU, 128 GiB RAM, and 600 GB HDD storage. The nodes are connected with 10 Gbps Ethernet links but we modify their latencies and bandwidths for our experiments using Distem.

We use Grid'5000 to reserve the physical nodes and deploy the file system image to them. Then, we use Distem to create the virtual nodes, load Debian 9 file system images to them, and create network interfaces. We also create several virtual networks to isolate different edge groups and gateway nodes. Using Distem, we modify the links latency and bandwidth settings to simulate an edge setting and a cloud setting as shown in table \ref{table:setup-lat-bw}.

\begin{enumerate}
   \item \textbf{The cloud setting:} to simulate a centralized cloud setting, we assume a low-bandwidth (100 Mbps) and high-latency (50 ms) link between the client and storage nodes, and high-bandwidth (1000 Mbps), very-low latency (in nanoseconds) links between the other nodes. This high latency simulates the typically large distance between clients and a data center. We carefully chose such parameters based on practical measurements of public cloud providers (e.g, Amazon EC2 and Microsoft Azure) average response times. Similar latency and bandwidth settings were used in \cite{mmog-edge} but we assume a simpler network model and smaller distances between gateway nodes. 
    
    We should note that cloud response time greatly depends on the client location. While clients closer to a data center will see lower response times, clients located farther away can have up to an order of magnitude higher response times. Also, clients outside North America and Europe generally get higher response times due to less efficient infrastructures and for the lack of enough data centers in other continents \cite{google-dc,amazon-dc,azure-dc}. \cite{game-measurement} reports that only 70\% of users in different US locations get an average of 80 ms response time or less (40 ms link latency) from the Amazon EC2 cloud. We set a 50 ms link latency instead for a more general assumption. We also use the online WAN latency Estimator tool \cite{wan-lat-estimator} for a lower-bound estimation of different link latencies since it only calculates propagation delay in fiber links. 

    \item \textbf{The edge setting:} simulates a typical edge deployment, with distances between hundreds of meters to a few kilometers. We simulate this with a latency of 2 ms between edge nodes, a latency of 5 ms between client and edge nodes in the same edge group, and higher bandwidth links in edge groups than farther-away gateway nodes. Unlike the cloud remote data centers, edge servers are assumed to be widespread to be within small distances from target clients and from neighboring edge servers (typically tens to hundreds of meters). Therefore, the links between such entities have low latencies. Based on the latencies required for optimal system performance, the edge nodes can be distributed to maximize the coverage while maintaining a manageable cost. However, solving such an optimization problem for node placement is outside the scope of this paper.
\end{enumerate}

\begin{table*}[t]
\begin{center}
 \caption{\label{table:setup-lat-bw} Link specifications to simulate edge and cloud settings. The following notation is used: Cli: client, St: storage node, Gw: gateway node.}
 \bgroup
\def\arraystretch{1.3}
\begin{tabularx}{\textwidth}
{
| >{\centering\arraybackslash}X
| >{\centering\arraybackslash}X
| >{\centering\arraybackslash}X
| >{\centering\arraybackslash}X
| >{\centering\arraybackslash}X
|
}
\hline
& \multicolumn{2}{|c|}{Edge} & \multicolumn{2}{|c|}{Cloud}\\
\hline
Link & Latency (ms) & Bandwidth (Mbps) & Latency (ms) & Bandwidth (Mbps)\\
\hline
Cli - St & 5 & 100 & 50 & 100 \\ 
\hline
St - St & 2 & 1000 & 0.05 & 1000 \\ 
\hline
St - Gw & 2 & 750 & 0.05 & 1000 \\ 
\hline
Gw - Gw & 10 & 500 & 0.05 & 1000 \\ 
\hline
\end{tabularx}
\egroup
\end{center}
\end{table*}

Unless specified otherwise, all experiments run 10,000 operations on each client in parallel, and the average results over all clients and operations are reported. We perform experiments in both the edge and cloud settings.
 
\subsection{Performance metrics}
We evaluate \Mechanism using multiple performance metrics to analyze its efficiency from different aspects. Below, we discuss each of the used metrics and their value in such an edge system.

\subsubsection{Operation response time}
The operation response time is the duration, measured from the client side, between sending a request to the storage node and receiving the response at the client side. This may include latencies caused by the network, consensus protocol, disk access, and possibly DHT routing. The type of operation specifies what the response time includes as follows:

\textbf{Read vs write requests}

Considering a single storage group (i.e., Raft group), each write operation requires replicating the data to a majority of servers in the group, according to the Raft protocol. Thus, a response time of a write operation includes at least the consensus protocol latency overhead for reaching a quorum, the communication latency between group members, and the overhead for writing to disk.

For read operations, there are two options. Linearizable reads are similar to writes in that each operation requires reaching a quorum with a majority of nodes to ensure data returned is the most recent one. On the other hand, serializable reads are more lightweight operations where any member node can directly return the result to the client without a quorum. This may result in returning stale data which may not be accepted in some cases. However, the quick response can improve performance a lot for non-sensitive data. etcd supports both linearizable and serializable reads but we use linearizable reads in our evaluations to analyze the system performance in the most demanding scenarios without making assumptions about the application. Applications using serializable reads will gain even more benefit from using \MechanismNoSpace. 

\textbf{Local vs global requests}

Local requests (requests to access local data) are returned directly from the same edge group to which the client sends their requests (their local edge group). Global requests (requests accessing global data), on the other hand, may involve some DHT routing overhead. When an edge node receives an access request for a global key-value pair, it first checks with its assigned gateway node if the key belongs to this edge group. If so, the key access is performed directly on that edge group. Otherwise, the DHT ring is traversed according to the Chord protocol to find the edge group responsible for the key. The request is forwarded to that edge group to perform the required operation. Thus, a global request may involve an additional overhead of DHT routing and edge-to-gateway and gateway-to-gateway communication.

\subsubsection{Throughput}
In addition to the response times, we also measure the system throughput defined as the number of operations the system can successfully complete in a second. The throughput is also measured from the client-side. In experiments where multiple client threads are used, the requests are distributed over the clients, and the reported throughput is the average responses from all the threads. For multiple clients communicating with different edge groups, we calculate the average throughput for each group's client, and take the average over all the clients.

\section{Performance Results} \label{sec-perf-resu}
In this section, we present the performance analysis results of \Mechanism using the emulation setup and performance metrics discussed in section \ref{section:emulation-setup}. The performed experiments analyze the efficiency and scalability of the system under different workloads and different configurations. Also, a complexity analysis of \Mechanism is included, followed by a discussion on energy consumption.

\subsection{Data-locality effect on performance}
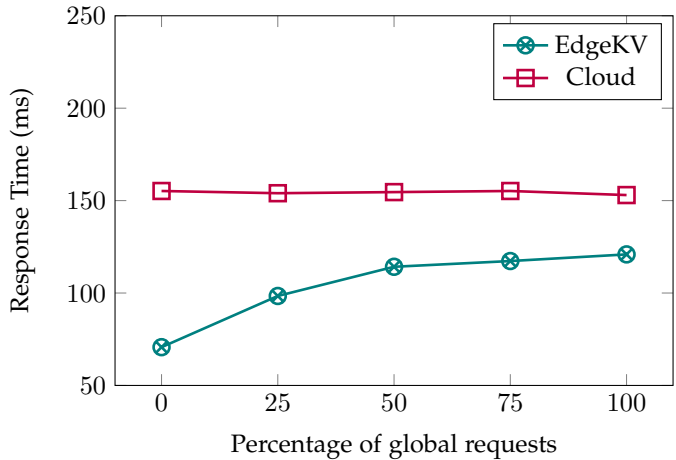
\begin{figure}[ht]
\centering
\begin{tikzpicture}
\begin{axis}[
    xlabel={Percentage of global requests},
    ylabel={Response Time (ms)},
    ymin=50, ymax=250,
    xtick={0,25,50,75,100},
    grid style=dashed,
    width = 9cm,
    height = 6.5cm,
    ]

    \addplot[
    color=black!00!teal,
    mark=otimes,
    line width=1pt,
    mark size=3pt
    ]
    coordinates {
    (0,70.7)(25,98.4)(50,114.2)(75,117.3)(100,120.9)
    }; 

    \addplot[
    color=black!00!purple,
    mark=square,
    line width=1pt,
    mark size=3pt,
    ]
    coordinates {
    (0,155.2)(25,154)(50,154.6)(75,155.2)(100,153)
    }; 
    \legend{EdgeKV, Cloud}
 
\end{axis}
\end{tikzpicture}
    \caption{Average write response time change with the percentage of global data in the requests.}    \label{fig:glob-perc-lat}
\end{figure}

\begin{figure}[h!]
\centering
\begin{tikzpicture}
\begin{axis}[
    xlabel={Percentage of global requests},
    ylabel={Throughput (ops/s)},
    xtick={0,25,50,75,100},
    ymajorgrids=true,
    grid style=dashed,
    width = 9cm,
    height = 6.5cm,
    ]

    \addplot[
    color=black!00!teal,
    mark=otimes,
    line width=1pt,
    mark size=3pt
    ]
    coordinates {
        (0,557.5)(25,412.1)(50,355.8)(75,356.9)(100,350.9)
    }; 

    \addplot[
    color=white!00!purple,
    mark=square,
    line width=1pt,
    mark size=3pt,
    ]
    coordinates {
    (0,296.8)(25,296.8)(50,298.1)(75,292.8)(100,296.8)
    }; 
    \legend{EdgeKV, Cloud}

\end{axis}
\end{tikzpicture}
    \caption{Average write throughput change with the percentage of global data in the requests.}    \label{fig:glob-perc-thr}
\end{figure}
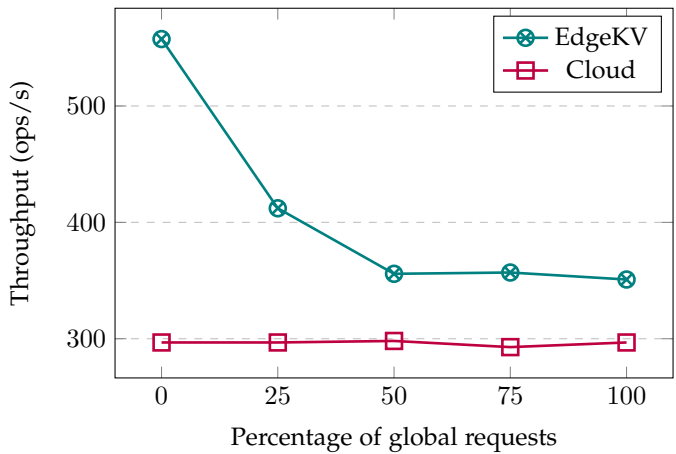

While a typical realistic edge use case would perform more local data access than global data access (e.g., VR and autonomous driving applications), we evaluate our system under loads with different global request percentages, as shown in figures \ref{fig:glob-perc-lat}, \ref{fig:glob-perc-thr}. We report the write latency and throughput results but the read operation results have an almost identical pattern. Both figures show that the edge setting outperforms the cloud one in both latency and throughput performance. It is interesting to notice that the change between 50\% - 100\% global data in throughput and latency is minimal. However, the performance decrease is high when the percentage of global requests is increased from 0\% to 50\%. Nevertheless, the edge still manages to keep its precedence over the cloud with 26\% lower latency and 19\% higher throughput at 50\% global write requests.

These results suggest with a low portion of the requests accessing global data, as in many practical applications, the system shows significantly better performance. This means the system is suitable for real-life scenarios but it also means careful care must be taken when designing applications targeted for the edge to minimize the number of global requests needed as much as possible.

\subsection{Request distribution effect on performance}
Requests in practical applications often do not follow a uniform distribution (where keys have equal popularity in a key-value store). Instead, a small set of the keys are often the most popular, known as the "hotset" or "hotspot" which the majority of read operations target. For example, in a social media application, a small portion of user accounts or posts are the most popular. Another common real-life request distribution pattern is the "latest" pattern where recently-inserted keys are generally more popular than older ones. This is common in systems that use sensor-collected information such as autonomous driving and VR applications as the recent information have higher value than older ones. 

\begin{figure}[ht]
    \centering
\begin{tikzpicture}
\begin{axis}[
    ybar,
    bar width = 5pt,
    width = 9cm,
    height = 6.5cm,
    enlargelimits=0.15,
    legend style={at={(0.5,0.95)},
    anchor=north,legend columns=-1,nodes={scale=0.9, transform shape}},
	ylabel={Response Time (ms)},
	symbolic x coords={latest, uniform, zipfian},
    xtick=data,
    xlabel=Request distribution,
    ymajorgrids=true,
    grid style=dashed,
    ]
\addplot[black!60!teal,fill=white!00!teal] coordinates {(latest,107.9) (uniform,115) (zipfian, 114.2)}; 
\label{edge} \addlegendentry{EdgeKV}

\addplot[black!60!purple,fill=white!00!purple] coordinates {(latest,149.6) (uniform,147.6) (zipfian, 154.6)}; 
\label{cloud} \addlegendentry{Cloud}

\end{axis}
\end{tikzpicture}
\caption{Average response time performance for update operations with 50\% of the requests accessing global data.} \label{fig:req-dist-latency}

\end{figure}
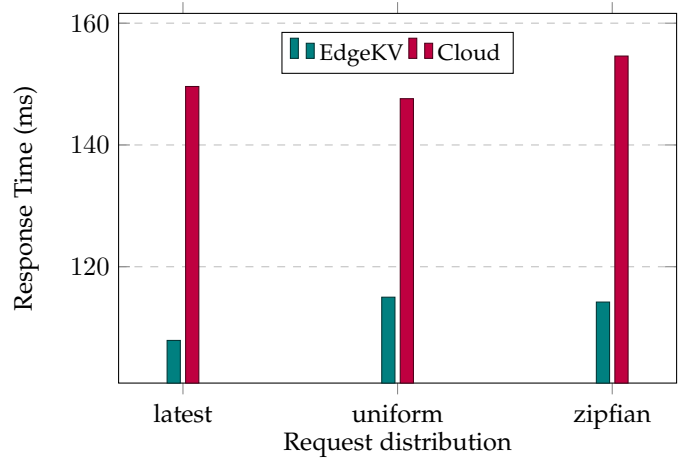

Since the performance of a key-value store typically changes based on the request distribution pattern, we evaluate \Mechanism with the three discussed popular patterns, namely, uniform, zipfian, and latest patterns. For the zipfian request distribution, we set the hotset size to 20\% of the total data and the percentage of operations that access the hot set to 80\%. YCSB chooses the keys for the hotset randomly from the total data inserted. 

\begin{figure}[h!]
    \centering
\begin{tikzpicture}
\begin{axis}[
    ybar,
    bar width = 5pt,
    width = 9cm,
    height = 6.5cm,
    enlargelimits=0.15,
    legend style={at={(0.5,0.95)},
    anchor=north,legend columns=-1,nodes={scale=0.9, transform shape}},
	ylabel={Throughput (op/s)},
	symbolic x coords={latest, uniform, zipfian},
    xtick=data,
    xlabel=Request distribution,
    ymajorgrids=true,
    grid style=dashed,
    ]
\addplot[black!60!teal,fill=white!00!teal] coordinates {(latest,378.8) (uniform,355.2) (zipfian, 355.8)}; 
\label{edge} \addlegendentry{EdgeKV}

\addplot[black!60!purple,fill=white!00!purple] coordinates {(latest,296.9) (uniform,308.8) (zipfian, 298.1)}; 
\label{cloud} \addlegendentry{Cloud}

\end{axis}
\end{tikzpicture}
\caption{Throughput performance for update operations with 50\% of the requests accessing global data.}\label{fig:req-dist-throughput}
\end{figure}
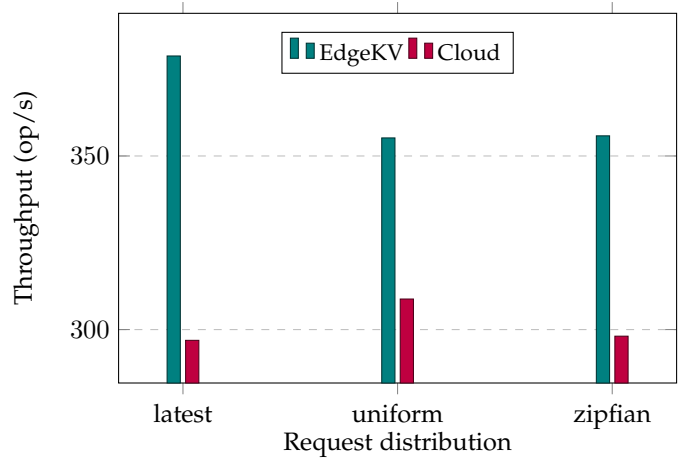

In figures \ref{fig:req-dist-latency} and \ref{fig:req-dist-throughput}, we can see that \Mechanism in an edge setting outperforms the cloud setting with a large difference, with all three request distribution patterns. Specifically, Fig. \ref{fig:req-dist-latency} shows that the write latency of the edge setting is 22\% - 28\% lower than the cloud setting with different request distributions. Similarly, in Fig.  \ref{fig:req-dist-throughput}, the edge setting outperforms the cloud one with 15\% - 28\% higher throughput. In both latency and throughput terms, \Mechanism achieves the best performance with the latest request distribution.

\subsection{Scalability: number of clients} 
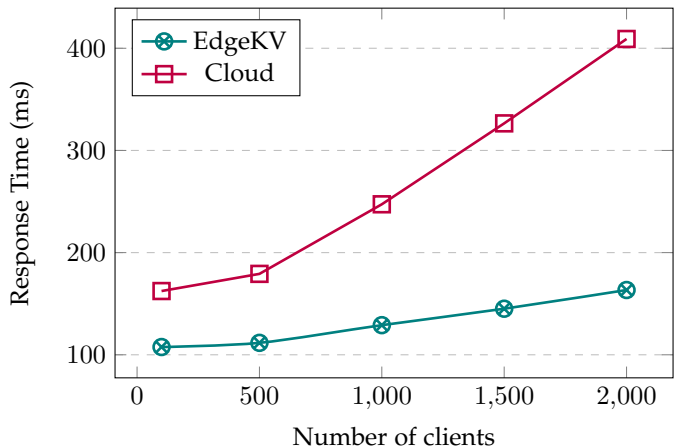
\begin{figure}[h!]
\centering
\begin{tikzpicture}
\begin{axis}[
    xlabel={Number of clients},
    ylabel={Response Time (ms)},
    legend pos=north west,
    ymajorgrids=true,
    grid style=dashed,
    width = 9cm,
    height = 6.5cm,
    ]
    \addplot[
    smooth,
    color=black!00!teal,
    mark=otimes,
    line width=1pt,
    mark size=3pt
    ]
    coordinates {
    (100,107.6)(500,111.7)(1000,129)(1500,145.1)(2000,163.4)
    }; 
\addplot[
    color=black!00!purple,
    mark=square,
    line width=1pt,
    mark size=3pt,
    ]
    coordinates {
    (100,162.4)(500,179.2)(1000,247.3)(1500,326.5)(2000,409.1)
    }; 
    \legend{EdgeKV, Cloud}
 
\end{axis}
\end{tikzpicture}
    \caption{Write response time scalability with the number of clients, using local requests only.}    \label{fig:etcd-cli-lat}
\end{figure}

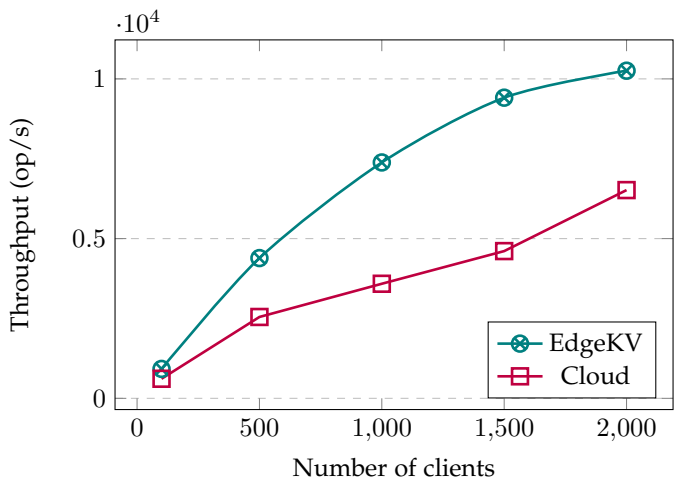
\begin{figure}[h!]
\centering
\begin{tikzpicture}
\begin{axis}[
    xlabel={Number of clients},
    ylabel={Throughput (op/s)},
    legend pos=south east,
    ymajorgrids=true,
    grid style=dashed,
    width = 9cm,
    height = 6.5cm,
    ]
    \addplot[
    smooth,
    color=black!00!teal,
    mark=otimes,
    line width=1pt,
    mark size=3pt
    ]
    coordinates {
    (100,921.1)(500,4390.3)(1000,7384)(1500,9411.7)(2000,10256.1)
    }; 
\addplot[
    color=black!00!purple,
    mark=square,
    line width=1pt,
    mark size=3pt,
    ]
    coordinates {
    (100,612.8)(500,2547)(1000,3586.8)(1500,4609.1)(2000,6518.8)
    }; 
    \legend{EdgeKV, Cloud}
 
\end{axis}
\end{tikzpicture}
    \caption{Write throughput scalability with the number of clients, using local requests only.}    \label{fig:etcd-cli-thr}
\end{figure}

with local requests
Here, we consider the performance of \Mechanism with local requests only by analyzing a single group of storage (etcd) nodes. We measure the response time and throughput of write operations from a client node in both the edge and cloud settings. Since the requests only access local data, the response time here is mainly the consensus latency (to reach a quorum among group members), including writing to disk, and communication with the client latency. In figures \ref{fig:etcd-cli-lat} and \ref{fig:etcd-cli-thr}, we show the results of these experiments. Specifically, we show how the write operations latency and throughput change with increasing the number of clients. The figures show that the edge setting achieves 1.5x - 2x higher throughput and 34 \% - 60\% lower latency than the cloud setting with different number of clients. 

The local requests performance results clearly show the advantage of utilizing \Mechanism in the edge setting. The close proximity of edge nodes to the clients allows for latency and throughput values not achievable by the remote cloud. Again, we see proof that defining the data locality in an application design can make a big difference in its performance.

\subsection{Scalability: number of clients with global requests}
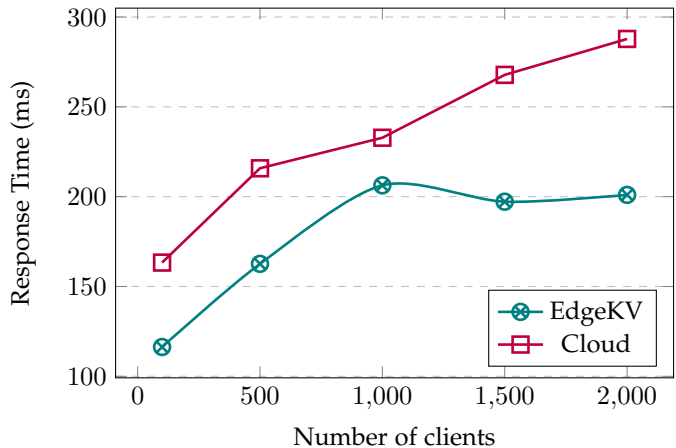
\begin{figure}[h!]
\centering
\begin{tikzpicture}
\begin{axis}[
    xlabel={Number of clients},
    ylabel={Response Time (ms)},
    legend pos=south east,
    ymajorgrids=true,
    grid style=dashed,
    width = 9cm,
    height = 6.5cm,
    ]
    \addplot[
    smooth,
    color=black!00!teal,
    mark=otimes,
    line width=1pt,
    mark size=3pt
    ]
    coordinates {
    (100,116.3)(500,162.6)(1000,206.3)(1500,197.2)(2000,201)
    }; 
\addplot[
    color=black!00!purple,
    mark=square,
    line width=1pt,
    mark size=3pt,
    ]
    coordinates {
    (100,163.3)(500,215.8)(1000,232.8)(1500,267.8)(2000,287.8)
    }; 
    \legend{EdgeKV, Cloud}
 
\end{axis}
\end{tikzpicture}
    \caption{Write response time scalability with the number of clients, using 50\% global requests}    \label{fig:cli-latency}
\end{figure}

\begin{figure}[h!]
\centering
\begin{tikzpicture}
\begin{axis}[
    xlabel={Number of clients},
    ylabel={Throughput (op/s)},
    legend pos=south east,
    ymajorgrids=true,
    grid style=dashed,
    width = 9cm,
    height = 6.5cm,
    ]
    \addplot[
    smooth,
    color=black!00!teal,
    mark=otimes,
    line width=1pt,
    mark size=3pt
    ]
    coordinates {
    (100,346.6)(500,1088.7)(1000,1853.5)(1500,2268.7)(2000,3128.2)
    }; 
\addplot[
    color=black!00!purple,
    mark=square,
    line width=1pt,
    mark size=3pt,
    ]
    coordinates {
    (100,278.6)(500,940.9)(1000,1637.2)(1500,2027.3)(2000,2493.2)
    }; 
    \legend{EdgeKV, Cloud}
 
\end{axis}
\end{tikzpicture}
    \caption{Write throughput scalability with the number of clients, using 50\% global requests}    \label{fig:cli-throughput}
\end{figure}
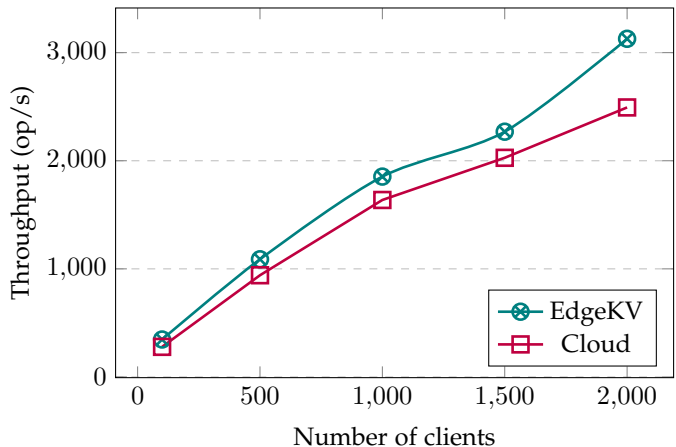

In this evaluation, we consider both local and global requests, by using workloads with 50\% global requests under the same configurations for both the edge and cloud settings. Figures \ref{fig:cli-throughput} and \ref{fig:cli-latency} show how the system scales almost linearly with the number of clients in both settings. While the edge throughput is slightly higher than that of the cloud (Fig. \ref{fig:cli-throughput}), the difference in response time change is considerable (Fig. \ref{fig:cli-latency}). Note, especially, when the number of clients increases from 1,000 to 2,000 clients, how the cloud average response time increases by about 24\% while the edge one is almost constant.

Comparing these results to the ones discussed in the previous section in figures \ref{fig:etcd-cli-lat} and \ref{fig:etcd-cli-thr}, we see that with a high number of clients, the effects of global requests on performance is higher (up to a factor of 4x higher throughput with the edge). Nevertheless, the edge setting keeps its advantage over the cloud in both cases.

\subsection{Request rate scalability with global requests}
\begin{figure}[h!]
\centering
\begin{tikzpicture}
\begin{axis}[
    xlabel={Request rate (op/s)},
    ylabel={Latency (ms)},
    legend pos=south east,
    ymajorgrids=true,
    grid style=dashed,
    width = 9cm,
    height = 6.5cm,
    ]
    \addplot[
    smooth,
    color=white!00!teal,
    mark=otimes,
    line width=1pt,
    mark size=3pt
    ]
    coordinates {
    (100,62.4)(200,74.7)(300,83.1)(400,87.5)(500,103.8)(600,110.8)
    }; 
\addplot[
    color=white!00!purple,
    mark=square,
    line width=1pt,
    mark size=3pt,
    ]
    coordinates {
    (100,121.6)(200,131.2)(300,147.5)(400,151.1)(500,156.3)(600,165.9)
    }; 
    \legend{EdgeKV, Cloud}
 
\end{axis}
\end{tikzpicture}
    \caption{Latency change with increasing request rates when 50\% of the requests are global requests}    \label{fig:latency-throughput}
\end{figure}
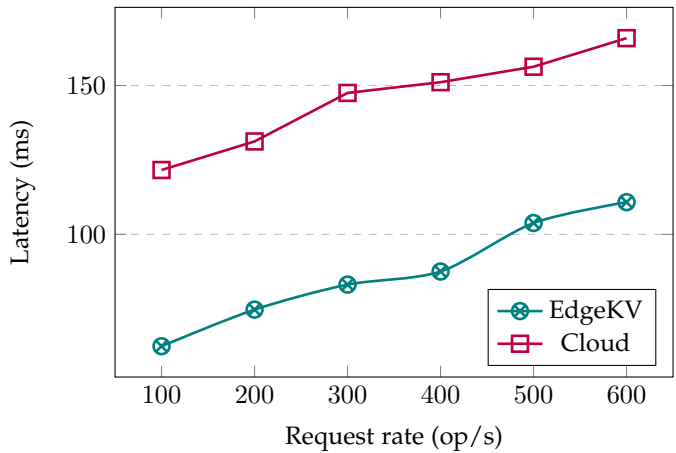

In this experiment, we use different request rates to analyze how much the system performance degrades with increasing request rates. We compare the performance of the system with 50\% global requests in both cloud and edge settings, using a 100 client threads, as shown in Fig. \ref{fig:latency-throughput}. The results show that in both settings the latency grows linearly with the request rate. Also, it is visible that there is a consistently large gap between the edge and cloud latencies, where the edge has, on average, 42\% lower latency than the cloud.

\subsection{Complexity Analysis of \Mechanism}
We provide a complexity analysis of \MechanismNoSpace's different operations in this section. Since the read operations discussed are linearizable reads (i.e., they require a consensus quorum to ensure data is up-to-date), the time complexity for both read and write operations are the same. However, the time complexity changes based on the data type:
\begin{enumerate}
    \item \textbf{Local data access} does not involve communication with gateway nodes or usage of the DHT overlay. Instead, local data is directly accessed from the edge group they were requested. Still, for fault-tolerance, a consensus quorum among edge nodes in the group is required for both read and write operations to ensure strong consistency. Assuming an edge group with n nodes, a majority of at least n/2 nodes is required to achieve consensus.
   \begin{center}
   Time complexity of local data access: $O(n)$.
   \end{center}    
    \item \textbf{Global data access}, on the other hand, may require lookup through the gateway nodes. If the key to access is in the responsibility of the edge group where the request is sent, the operation is performed directly in that group, achieving the same time complexity as local data. However, if another edge group is responsible for the key, then the responsible group needs to be decided first through a lookup operation in the DHT overlay. A DHT achieves lookup in O(log(m)) with m gateway (i.e., DHT) nodes. After the responsible gateway node is found (hence its corresponding edge group), a consensus quorum needs to be achieved between the group nodes in O(n) time for the data access. 
    \begin{center}
    Time complexity of remote global data access:
    $O(n + log(m))$
    \end{center}
\end{enumerate}

Space complexity on each edge node is computed as follows: Each edge node has a copy of all the local data stored in the edge group, which is O(L*S) assuming L is the number of keys in the local group and S is the average size of a local key-value pair. Besides, since the global data are uniformly distributed over the edge groups, each group also gets global storage of O(G*T/m) where G is the total number of global keys in the system, T is the average size of a global key-value pair, and m is the number of edge groups (i.e., number of gateway nodes).

\begin{center}
Space complexity of a single edge node:
$O(L*S + G*T/m)$
\end{center}

On the other hand, gateway nodes do not persist any key-value pairs. They are only required to store the finger tables for lookup through the DHT overlay. So, for m gateway nodes:

\begin{center}
Space complexity of a single gateway node: $O(log(m))$
\end{center}

\subsection{Energy Considerations in the Edge}
Considering the energy consumption in distributed Peer-to-Peer (P2P) system design, developing mechanisms that improve energy efficiency \cite{2016p2psurvey}, energy cost models \cite{fgcs14}, and metrics of energy efficiency \cite{resdn2020} are significant. Communication among edge nodes can deliver services to peers with minimal reliance on the cloud through resource and capability sharing of cooperative fogs. Unfortunately, this has huge drawback on edge nodes power, and currently the most effective model is through cooperation among the edges \cite{al2019profitable}. In a centralized cloud setting, each request needs to be sent to the remote cloud, through high latency links, which can cause the cloud servers and network links to the cloud to be overloaded, wasting both time and bandwidth. Besides, in the cloud, all application data is treated equally (i.e., giving the same performance guarantees) regardless of the location of the source and consumers of data and the number of consumers. On the other hand, \Mechanism considers the location context of users and the fact that some data may need to be shared only with a small number of users in the same geographical area, namely local data. This allows some of the data to be stored in the proximity of its users, saving bandwidth and providing low-latency access while maintaining consistency. Therefore, the use of local data in \Mechanism achieves higher energy saving than the data-type-agnostic cloud, especially since in many applications such as autonomous driving and VR applications, local data is more frequent than global data \cite{mmog-edge, safara2020prinergy}. Moreover, since global data is distributed uniformly over multiple edge groups in different locations, the resource usage per each edge group becomes smaller as the system grows larger, making resources less likely to be overloaded.


\section{Practical Requirements for Scalability}\label{sec-prac-sca}
\subsection{Virtual Nodes}
To ensure a more uniform load-balancing between the edge groups for global data, the concept of \textit{virtual nodes} could be utilized in the overlay. For each physical gateway node, multiple virtual nodes (e.g., log(N) virtual nodes where N is the number of physical gateway nodes) can be assigned on the overlay. This has been shown to significantly improve load balancing with the cost of increasing storage space for routing information. However, such storage is insignificant in practice. Virtual nodes can be especially useful when different edge groups have varying resources. More virtual nodes could be assigned to the gateway nodes associated with the more powerful edge groups. This means that such groups will store a bigger portion of the global data than other, less powerful, groups.

\subsection{Caching}
With a large-scale deployment of \MechanismNoSpace, the average distance between a client and a random remote edge group will get larger, causing higher average latency for global data access. To improve the access latency, especially for frequently accessed remote data, caching can be utilized. Each edge node in the system can cache some of the global data that is stored in other edge groups. Choice of which data to cache depends on the application requirements but strategies such as choosing the most recent data or the most recently accessed data can be applied. The size of the cache can grow or shrink according to the available free storage on the edge node in a specific edge group. We note that to ensure strong consistency (i.e., linearizable reads), reading cached global data would still involve contacting a remote node to validate the cache is up-to-date. However, if reading stale values is tolerated (i.e., serializable reads), then the cached value would be directly returned without contacting the remote group.

Since finding a key's location on the DHT overlay has O(log(n)) time complexity, caching in the gateway node can also be useful. The locations of popular or recent keys can be added to the gateway cache to avoid the key lookup overhead. The location information would contain the responsible gateway node's identifier in the overlay and its network address (e.g., IP address and port number).

\subsection{Inter-group fault tolerance}
A global key-value pair is replicated on each edge node in its assigned group (similar to local key-value pairs), so it would still be accessible even if a minority of the group members fail. However, if a majority of the members fail, or if the entire edge group becomes inaccessible to other parts of the overlay (e.g., because of network partitions or link congestion), then the global data stored at that group becomes unavailable to the rest of the system. Because of the distributed nature of the edge setting, network partitions are more common in the edge than in the cloud. To solve these issues, we propose the idea of a \textit{backup group}. For each edge group in the system, we assign another group as its backup group. This assignment would follow static rules so any node can identify the backup group given the original group identifier. A simple approach would be to specify the backup group as the first group directly following node in the overlay.

A backup group is kept up-to-date with the original group as follows: the backup group is included in the original group’s members list as a non-voting member. This means that it receives all consensus requests as the other members and is notified of the committed entries. However, it is not counted in the consensus majority. A backup group does not receive data access requests from other groups until the original group becomes unreachable. Even then, the backup group is used for read operations only. This is important to ensure that the states of original and backup groups will not diverge. The backup group may have stale data for some time, but it will still be possible to correct the state once the original group becomes available again. 

\section{Conclusion and Final Remarks}\label{sec-concl}
We introduced \MechanismNoSpace, a distributed storage solution for the edge to help application developers worry less about the underlying infrastructure and focus more on the application design. \Mechanism abstracts away the details of the edge infrastructure with well-defined interfaces and a location-transparent data placement strategy. Moreover, \Mechanism provides fault-tolerance and strong consistency guarantees through data replication in the local edge groups. Nevertheless, high scalability is achievable with an efficient DHT-based overlay to connect edge groups in different locations. The modular design of \Mechanism allows the easy replacement of any of the modules to provide different storage types, varying consistency and latency guarantees, or to introduce application-specific requirements.

We have implemented a prototype of \Mechanism in Golang and presented our performance analysis results from different aspects. \Mechanism achieves 26\% lower latency and 19\% higher throughput than a centralized cloud solution with 50\% global requests under the same testing conditions. We have also shown that \Mechanism scales better than the cloud with the number of requests with an average of 42\% lower latency even with 50\% of the requests accessing global data. Finally, we have demonstrated how the different request distribution patterns affect \MechanismNoSpace, and that \Mechanism performs 22\% - 28\%  faster writes on average with 15\% - 28\% higher throughput.

With its decentralized architecture, \Mechanism does not depend on the cloud to run. Nevertheless, it can still benefit from the cloud in a number of ways. For inter-group fault tolerance, backups could be uploaded to the cloud and restored later in case of network partitions or group failures. In addition, a central entity as the cloud can be used for the initial bootstrapping of the edge groups and the assignment of gateway nodes to edge groups.

\Mechanism is designed as a general-purpose strongly-consistent key-value store. However, due to its flexible and modular design, it can be used in different configurations to suit a multitude of use cases. As a future work, \Mechanism can be used with different replication techniques to provide weaker forms of consistency for lower response times. Alternatively, utilizing an adaptive consistency strategy based on the criticality of data \cite{adaptive-consistency} and the use of different DHT structures for replication \cite{pyramid20}\cite{hassanzadeh2018decentralized} per edge group can be investigated. Similarly, different storage drivers can be used for storing different data types. For example, the key-value store can be replaced with a relational SQL database or with a NoSQL data store. \Mechanism performance could be further improved by using low-latency persistent storage such as the Non-Volatile Memory (NVMe) and low-latency communication transport such as Remote Direct Memory Access (RDMA). RDMA avoids the overhead of the traditional TCP transport by performing zero-copy transfer and directly accessing memory or non-volatile memory. A few RDMA-enabled consensus protocols already exists \cite{rdma-consensus} and could be used in the replication manager module to perform faster replication

\section*{Acknowledgements}\label{sec-ack}
    Deployment and experiments in this study are carried out using the Grid'5000 testbed, supported by a scientific interest group hosted by Inria and including CNRS, RENATER as well as other organizations (https://www.grid5000.fr). A very preliminary version of this study was submitted for conference presentation \cite{edgekv-short-paper}. This work was partially supported by the College of Engineering, Al Ain University, under Grant ERF-20.
    
    \balance
\bibliographystyle{IEEEtran} 
\bibliography{main}

\ifCLASSOPTIONcaptionsoff
  \newpage
\fi

\end{document}